\documentstyle[11pt,draft,epsf]{article}

\newcommand{\beq}{\begin{equation}}
\newcommand{\eeq}{\end{equation}}
\newcommand{\beqa}{\begin{eqnarray}}
\newcommand{\eeqa}{\end{eqnarray}}

\begin{document}

\begin{flushright}
INFNCA-TH9513\\
MPI-PhT/95-31 \\
hep-lat/9504004
\end{flushright}

\vfill

\begin{center}{\Large\bf
Monte Carlo simulations of
\\
\medskip 4d simplicial quantum gravity\footnote{
Contribution to the special issue of the Journal of Mathematical Physics
on {\em Quantum Geometry and Diffeomorphism-Invariant Quantum Field
Theory},  edited by Carlo Rovelli and Lee Smolin.}
}
\end{center}
\vfill

\begin{center}
\large Bernd Br\"ugmann
\end{center}

\begin{center}
{\em Max-Planck-Institut f\"ur Physik,
 \\ F\"ohringer Ring 6, 80805 M\"unchen, Germany \\}
{\tt bruegman@mppmu.mpg.de}
\end{center}

\smallskip

\begin{center}
\large Enzo Marinari
\end{center}

\begin{center}
  {\em Dipartimento di Fisica and Sezione Infn, Universit\`a di Cagliari,\\
  Via Ospedale 92, 09100 Cagliari, Italy \\}
  {\tt marinari@ca.infn.it}
\end{center}
\vfill

\begin{center}
\large\bf Abstract
\end{center}
\medskip

\noindent
Dynamical triangulations of four-dimensional Euclidean quantum gravity
give rise to an interesting, numerically accessible model of quantum
gravity.  We give a simple introduction to the model and discuss two
particularly important issues. One is that contrary to recent claims
there is strong analytical and numerical evidence for the existence of
an exponential bound that makes the partition function
well-defined. The other is that there may be an ambiguity in the
choice of the measure of the discrete model which could even lead to
the existence of different universality classes.

\vspace*{\fill}
\newpage

\section{Introduction}

Dynamically triangulated random sur\-faces (DTRS) \cite{DYTRAS} play an
important role in the efforts to develop a coherent description of quantum
gravity. The (euclidean) space-time is approximated by a $d$-dimensional
simplicial triangulation, where the link length is constant, equal to $1$,
but the connectivity matrix is a dynamical variable.

The most important advances have been obtained in two-dimensional
quantum gravity, where DTRS are simplicial triangulations of a $2d$
manifolds. The analytic success of matrix models, which can be for
example exactly solved in the case of pure $2d$ gravity \cite{EXACT},
has strongly encouraged this approach.  The results obtained in the
triangulated approach and in the continuum lead to consistent
predictions for correlation functions and critical exponents.

Dynamical triangulations are also potentially relevant in four
dimensions. One can hope that a sensible, non-perturbative definition
of the quantum gravity theory can be obtained in some {\em scaling}
limit of the theory of $4d$ hyper-tetrahedra. This approach has much
in common with Regge calculus, where the connectivity is fixed but the
functional integration runs over the link lengths. The underlying
principle is clearly very similar, and one could say that DTRS have
the status of an improved Regge calculus. The fact that the
coordination number can vary in the DTRS makes it easier to describe a
situation in which long spikes play an important role.

We face the usual problem inherent in discretizing a theory, i.e.\ the
discretization scheme can break some of the continuous symmetries,
which will have to be recovered in the continuum limit (if there is
one). Indeed, Wilson lattice gauge theories have taught us an
important lesson. The fact that gauge invariance is exactly conserved
in the lattice theory, for all values of the lattice spacing $a$, is
in that case crucial: it would have been very difficult to establish
firm numerical results if one would have had to care about the
presence of non gauge-invariant corrections, which would disappear only
in the $a \to 0$ limit. In the case of quantum gravity, diffeomorphism
invariance plays such a crucial role, and DTRS are diffeomorphism
invariant by construction, at least on the space of piecewise flat
manifolds. Hence part of the difficulties Regge calculus has in
forgetting about the lattice structure are eliminated a priori in the
DTRS lattice approach. The results of \cite{Regge} actually show that
the conventional application of Regge calculus to quantum gravity in
two dimensions fails to reproduce the analytical results, although a more
sophisticated approach may still succeed \cite{Regge}.

There are two more important points to stress. The first one is that
in the DTRS approach in $3d$ and $4d$, as opposed to $2d$, we can try
to make sense out of the pure Einstein action, without, for example,
curvature squared terms. Even though the partition function formally
diverges, at fixed volume the local curvature is bounded both from
below and from above. Therefore we can study the theory at fixed (or
better quasi-fixed, see later) volume, and look for the existence of a
stable fixed point in the large volume limit. A second order phase
transition with diverging correlation lengths, in the statistical
mechanics language, would allow us to define a continuum limit which
is universal and is not influenced by the details of the underlying
discrete lattice structure. Precisely this scenario constitutes one of
the best hopes we have to find a consistent quantum theory of gravity.
It could be a way to give a non-perturbative definition of
euclidean quantum gravity based on the Einstein action.

On the other hand, we believe that there are at least four potentially
deep issues in $4d$ simplicial quantum gravity on which the final
success of this model hinges. The first one is related to the
unrecognizability of $4$-manifolds, which may invalidate the whole Monte
Carlo approach.  The second one is the question whether there exists
an exponential bound for the partition function such that the model is
well-defined. This question is likely to have been settled in the
affirmative recently.  Third, it is quite unclear what role the
unknown measure of the path integral approach to quantum gravity plays
in the triangularized model, and it may well be that there are
different universality classes.  Finally, since Euclidean general
relativity cannot in general be extended to the physical, Lorentzian
sector, one may wonder what the appropriate physical observables in
the theory are.

Our contribution to this volume will be to exemplify concrete
numerical problems in the four-dimensional case.  In section 2, we
introduce the model and Monte Carlo simulations. We comment
on the issue of ergodicity and state what the simplest
observables are that we evaluate. In section 3, we introduce the problem
of the exponential bound and describe a detailed analysis of the
numerical simulations addressing this problem. In section 4, we present
some preliminary results about the universality structure for a
one-parameter family of measures that contains the trivial measure as
a special case, which is the one considered in all the other
investigations. We close with a summary in section 5.

\section{Simplicial quantum gravity in four dimensions}

\subsection{The Model}

The model we consider is based on the four-dimensional Euclidean
Einstein-Hilbert action,

\beq
S_E[g] = \lambda \int d^4x \sqrt{g} - \frac{1}{G} \int d^4x \sqrt{g} R(g),
\eeq
where $\lambda$ is the cosmological constant, $G$ is the gravitational
constant, $g$ is the determinant of the metric, and $R(g)$ its scalar
curvature.

As discussed in \cite{AgMi92,AmJu92}, if one considers only manifolds
which are simplicial complexes with $S^4$ topology and defines the
metric by the condition that all edges have length 1, then the volume
integral $V$ and the net scalar curvature $R$ can be replaced by

\beqa
V \equiv \int d^4x\sqrt{g} & \leftrightarrow & N_4[T], \\
R\equiv \int d^4x\sqrt{g} & \leftrightarrow & \frac{2\pi}{\alpha} N_2[T] - 10
N_4[T],
\label{defR}
\eeqa
where $N_i[T]$ denotes the number of $i$-simplices of the
triangulation $T$ and $\alpha$ is derived from the condition that for
approximately flat triangulations the curvature vanishes, $\alpha =
arccos(1/4) \approx 1.318$.  The action takes a very elegant and
simple form \cite{AgMi92,AmJu92},

\beq
S_E[T] = k_4 N_4[T] - k_2 N_2[T],
\label{se}
\eeq
where the coupling constants are

\beq
k_4 = \lambda+10/G, \quad k_2 = \frac{2\pi}{\alpha G}.
\eeq
Notice that (\ref{se}) is the most general action of the type
$S_E=\sum_iN_i$ in four dimensions. Euler's relation for $S^4$ and the
Dehn-Sommerville relations leave only two of $N_0,\ldots,N_4$
independent. The number of vertices $N_0$ in the simplicial complex
for example is

\beq
N_0 = \frac{1}{2} N_2 - N_4 + 2.
\eeq
In addition, there are inequalities between the $N_i$. Denote by $o(a)$
the order of vertex $a$, i.e. the number of four-simplices that contain
$a$. For the average order of simplices we have

\beq
5 \leq \frac{1}{N_0} \sum_a o(a) = \frac{5 N_4}{N_0} < \infty,
\eeq
which implies

\beq
2 < \frac{N_2}{N_4} < 4.
\eeq
Hence the average curvature is asymmetrically bounded from below
and above, $ -0.614 < R/N_4 < 12.0 $, which is a reflection of the fact
that the choice of $S^4$ topology restricts the choice of possible metrics.

The purpose of the above discretization is to make the path integral
well-defined,

\beq
\int {\cal D}g\, e^{-S_E[g]} \leftrightarrow \sum_{T} e^{-S_E[T]},
\label{conttodisc}
\eeq
where the integration over all metrics is replaced by a summation over
all triangulations with $S_4$ topology.
By itself, this replacement is not sufficient to define a finite
partition function, rather there are conditions on the coupling constants
as we will discuss in section 2. Notice also that
although triangulations allow us in principle to perform a summation over
different topologies, this sum is known to diverge badly.

In conclusion, we study a rather simple looking model for Euclidean quantum
gravity defined by a grand canonical partition function $Z(k_4,k_2)$ and
a canonical partition function $Z(N_4,k_2)$,
\beqa
Z(k_4,k_2) & = & \sum_{N_4} e^{-k_4N_4} Z(N_4,k_2),
\label{grand}
\\
Z(N_4,k_2) & = & \sum_{T:|T|=N_4} e^{k_2N_2[T]},
\label{canonical}
\eeqa
where we have split the sum over all triangulations of $S^4$ into a sum
over all possible volumes (equal to the number of 4-simplices $N_4$) and
a sum over all triangulations $T$ with volume $|T|$ equal to $N_4$.

The existence of a non-trivial continuum limit of the theory would
manifest itself as a critical point, with a diverging correlation
length and non-trivial critical exponents. A diverging correlation
length allows physical observables to forget about the original
discrete nature of the model. The issue of the existence of such a
limit, and of the features which would characterize such a continuum
theory is the main point of the discussion we are summarizing here.

\subsection{Monte Carlo simulations of the model}

The Monte Carlo evaluation of the partition function (\ref{grand}) is
largely standard once an ergodic random walk through the space of
triangulations has been defined (see e.g.  \cite{Br93} for a detailed
description).  Notice that while for dynamical triangulations the
action of gravity has become a very simple linear function of two
global numbers, the non-trivial part of the theory is represented by
the complexity of the space of triangulations.  This is reflected in
the ergodicity problem of the random walk: while there does exist a
simple set of five local moves that is ergodic in the space of all
triangulations of compact four-manifolds \cite{GrVa92}, it is not
finitely ergodic \cite{NaBe93}.

Let us first introduce these elementary moves and then comment on the
ergodicity problem. Denoting a four-simplex by its five vertices,
\beqa
abcde & \leftrightarrow & Aabcd, Aabde, Aacde, Abcde,
\label{m04}
\\
abcdA, abcdB & \leftrightarrow & ABabc, ABabd, ABacd,
\label{m13}
\\
abcAB, abcAC, abcBC & \rightarrow & ABCab, ABCac, ABCbc,
\label{m22}
\eeqa
where $a, b, \ldots$ and $A, B, \ldots$ are the vertices which are common
to all four-simplices on the left and right-hand sides, respectively.
Notice the regular structure which involves permuting all indices of one
type on one side keeping the others fixed and doing the opposite on the
other side. Move $i$ is the exchange of  $i$-simplices for appropriate
$(4-i)$-simplices. For example, move 0 as given by going from right to
left in (\ref{m04}) removes the vertex $A$ common to precisely five
4-simplices creating one new 4-simplex. The two dimensional analog of
(\ref{m04}) is adding and removing a vertex inside a triangle.

There are two restrictions on when a move may be performed which ensure
that the simplicial complex does not change topology. First of all, move
$i$ is only possible if the order of the simplex which is to be removed
is $5-i$. For move 0, for example, a vertex can only be removed if it is
of order 5, since otherwise its neighboring vertices do not form a
four-simplex which always has five vertices (of order 3 in two
dimensions). And second, a move is not allowed if it creates simplices
which are already present. For example, move 3 --- left to right in
(\ref{m13}) --- introduces an new link $AB$ that, if already part of the
simplicial complex, would lead to overlapping four-volumes.

The computer challenge posed by dynamical triangulations is to
implement a data structure for the simplicial complex which allows
efficient updating under the moves (\ref{m04})--(\ref{m22}), and which
in particular is not static (e.g.\ \cite{Br93}).  A novel approach to
speed up thermalization is called baby-universe surgery
\cite{chop}. This approach works very well in the elongated phase of
the theory (see section 2.3), showing its validity at the critical
point and in the crumpled phase might require further study.

Even though the elementary moves introduced above are ergodic, they
are not finitely ergodic \cite{NaBe93}.  This is directly related to a
result by Markov, stating that most simplicial four-manifolds are
not algorithmically distinguishable.  It is not known whether this is
the case for $S^4$.  There obvious might be a serious problem for the
Monte Carlo simulations if the random walk does not reach physically
relevant portions of the space of triangulations.  There are numerical
attempts to detect the non-distinguishability in Monte Carlo
simulations for $S^4$ without any indication of such \cite{AmJu2}, but
the same is true for $S^5$ which is known to be non-distinguishable
\cite{Ba94}.  Regarding these attempts it may well be that the systems
studied are far too small or the method of study is not appropriate.
The physical relevance of this is quite open. It could very well be
that the measure is strongly peaked close to {\em distinguishable}
manifolds, and that the problem does not arise with finite measure in
the continuum limit.

A technical problem related to the elementary moves is that no set of
elementary moves is known that leave $N_4$ invariant.
In the simulations fixed $N_4$ is approximated by allowing $N_4$ to vary
in a certain range which is a small fraction of $N_4$ but as large as
feasible to approach ergodicity.

\subsection{Basic results}

Of interest are the expectation values of observables depending a priori
on both $k_2$ and $k_4$.
When simulating the system described by the partition function
$Z(k_4,k_2)$ with variable volume, one finds that there exists a line
$k^c_4(k_2)$ in the plane of the coupling constants such that for any
$k_2$, if $k_4 > k_4^c$ the volume is driven towards zero, and if $k_4 <
k_4^c$, the volume goes to infinity. The larger the deviation from $k_4^c$
the faster the trend. For these reasons $k_4^c(k_2)$ is often called the
critical line (although it has nothing to do with statistical criticality).
Whether one can prove the existence of the critical line is the
subject of section 3.

A typical simulation is then performed for given $k_2$ and a fixed volume
$N_4^0$, and $k_4 = k_4^c(k_2) $ is
determined dynamically during the run. The same technical trick
that allows one to keep the volume near a given value stabilizes the
balance between drifting towards zero or infinite volume: one adds an
artificial `potential well' to the action that is dynamically adjusted to
be centered at $k_4^c$.

On the critical line one then looks for a second order phase
transition, since this is where a continuum theory might be
defined. Indeed, the data are consistent with a continuous phase
transition near $k_2^c \approx 1.1$. In \cite{continuous} the claim is
made that the smooth nature of the transition has been proved. One
distinguishes two phases of the model, the elongated phase where the
Haussdorff dimension approaches 2, and the crumpled phase where the
Haussdorff dimension approaches infinity.

How to measure this dimension is somewhat tricky, but it would
be directly related to an effective physical dimension, e.g. like the one
experienced by a test particle. One of the appealing features of
simplicial quantum gravity is that in this way one could derive the
number of physical dimensions, and in fact at the critical point the
dimension is observed to be close to four (e.g.\ 4.6 in \cite{AgMi93} for the
diffusion equation of a heavy test particle).

Finding good physical observables in simplicial quantum gravity is an
important problem. Notice that because the quantized metric field is
not an observable of quantum gravity (because it is not diffeomorphism
invariant), one cannot sensibly study correlation functions based on
the configuration variables themselves. Integration of a density over
the manifold leads to a diffeomorphism invariant quantity, which is
the case for example for the total curvature $R$, (\ref{defR}). In
integrated form one can also study objects like the
curvature-curvature correlation \cite{continuous,xxx}.

There are many interesting observations about simplicial quantum gravity
that could be discussed at this point, but let us now focus on two
particular topics as promised in the introduction.

\section{Is the partition function well-defined?}

As we have explained in section 2, there are many indications that the
partition function $Z(k_4, k_2)$ defines a sensible discrete model for
quantum gravity.  However, in \cite{CaKoRe94} Catterall, Kogut and Renken
put forward the claim that the partition function is actually ill-defined
since based on their new numerical data it is not exponentially bounded in
the large volume limit.  This prompted a more careful examination of the
partition function, both numerically and analytically, with the net outcome
that an exponential is very likely to exist after all.

Let us pose the problem.  Finiteness of the partition function $Z(k_4,k_2)$
defined in (\ref{grand}) is related to the existence of an exponential
bound for $Z(N_4,k_2)$ as follows.  Suppose there exists an exponential
bound for the canonical partition function,

\beq
   Z(N_4,k_2) \sim e^{k_4^c(k_2) N_4}\ ,
\eeq

\noindent for large $N_4$ and some constant $k_4^c(k_2)$.
Then the partition function $Z(k_4,k_2)$ is finite for $k_4 > k_4^c(k_2)$
and divergent for $k_4 \le k_4^c(k_2)$.

The question of the existence of an exponential bound for the canonical
partition function is directly related to the asymptotic behavior of the
number of triangulations for a given volume, ${\cal N}(N_4)$, which might
grow as fast as $(5N_4)!$.
Since $2 N_4 < N_2 < 4 N_4$,

\beqa
    {\cal N}(N_4) \le Z(N_4,k_2) \le e^{4 k_2 N_4} {\cal N}(N_4)
    &&  \quad \mbox{if $k_2 \ge 0$,}
\\
    Z(N_4,k_2) < e^{2 k_2 N_4} {\cal N}(N_4) < {\cal N}(N_4)
    &&    \quad \mbox{if $k_2 < 0$.}
\eeqa

\noindent Hence the
existence of an exponential bound on ${\cal N}(N_4)$ implies the same for
the canonical partition function for arbitrary $k_2$, and if there exists
an exponential bound on the canonical partition for a single value $k_2 \ge
0$ then it exists for all $k_2$.

To summarize what is known analytically,
in two dimensions there is the classical result of \cite{Tu62} giving an
explicit exponential bound on the number of triangulations. In three
dimensions, there does not yet exist a proof and the situation is quite
subtle, e.g. \cite{DuJo}. In \cite{BaBrCaMa}, a novel method for counting
minimal geodesic ball coverings of two- and four-dimensional manifolds is
developed, which is based on certain finiteness theorems about the number
of homeomorphism types for geometrically bounded manifolds. While there are
strong plausibility arguments that counting such balls gives also an
upper bound on the number of triangulations, the issue is not completely
settled. For example, the construction does not apply directly to
piecewise-linear manifolds as we are considering here, although a
heuristic connection can be made. But it seems likely that an analytical
proof of the existence of the exponential bound can be constructed.

\begin{figure}
\epsfxsize=400pt
\centerline{\epsffile{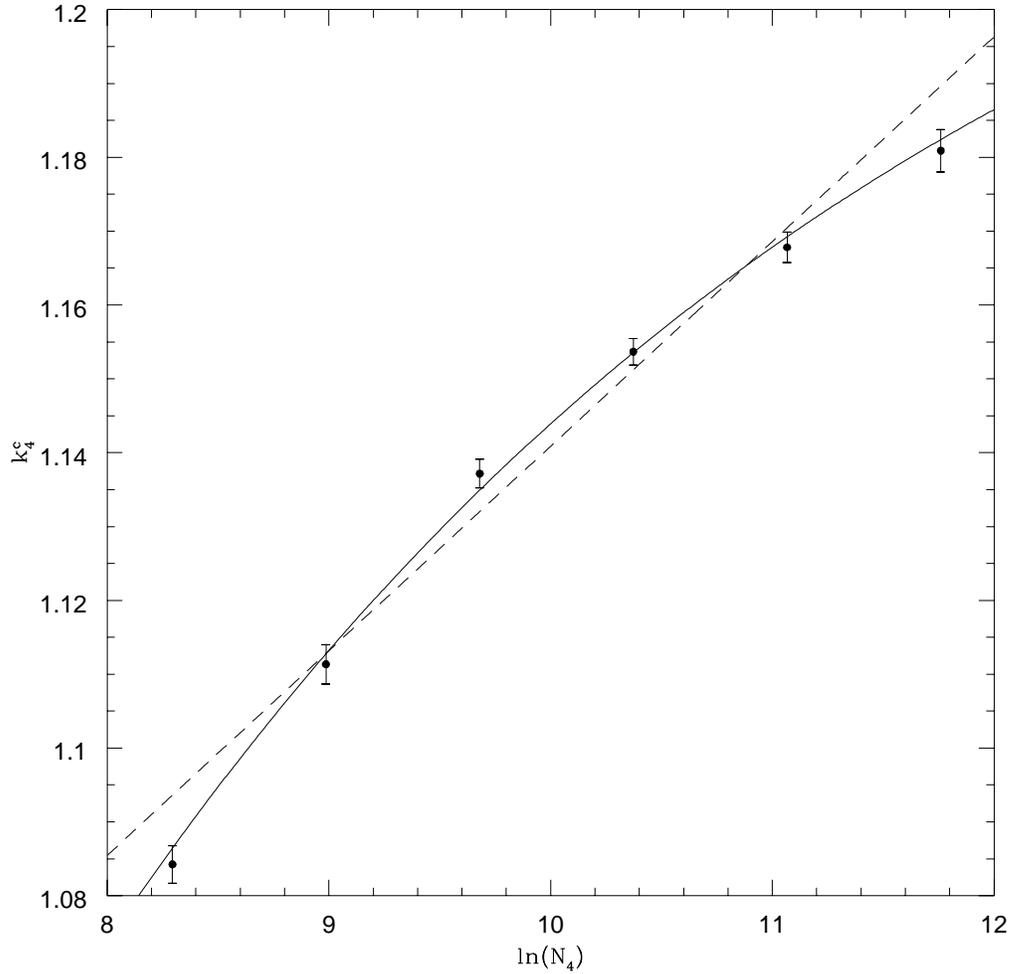}}
\caption[a]{\protect\label{figk4cN4} $k_4^c$ versus $\ln(N_4)$ for $k_2 =
0$.  Values for $N_4$ are $4000$, $8000$, $16000$, $32000$, $64000$ and
$128000$.  With the dashed line we give our best logarithmic fit, with the
solid line we give our best fit to a converging power, with $\alpha=.25$.
Both fits have two free parameters.  The $\chi^2$ of the power fit is ten
times better than the one of the logarithmic fit.  }
\end{figure}

On the numerical side, there is strong numerical evidence for the
existence of an exponential bound \cite{BrMa95}. As already mentioned,
in four dimensions the numerical data of the initial investigations
did not show any inconsistency related to the absence of an
exponential bound.  If an exponential bound $\exp a N_4$ to the
canonical partition function exists, then

\beq
  k_4^c(N_4) = a + b
  N_4^{-\alpha},
  \label{powerfit}
\eeq
where $N_4^{-\alpha}$ represents a natural polynomial correction to the
exponential. If instead of an exponential bound only a factorial bound
holds, then one expects

\beq
k_4^c(N_4) = a + b \log N_4.
\label{logfit}
\eeq
In figure (\ref{figk4cN4}) we show a typical plot of $k_4^c$ versus $\log
N_4$ up to $N_4 = 128k$ at $k_2 = 0$ based on \cite{BrMa95}.

In \cite{CaKoRe94}, data for $k_2 = 0, 0.25, 0.5$ were presented for
volumes up to 32k simplices, and since a logarithmic fit as in
(\ref{logfit}) seemed reasonable, absence of an exponential bound was
claimed.  In \cite{BaSm} and \cite{AmJu}, the accuracy of the data was
improved upon somewhat, but it was also noted that a small power in
(\ref{powerfit}) may fit the available data about as well as a
logarithm.  The authors of \cite{BaSm} prefer the logarithmic fit,
while in \cite{AmJu} for volumes up to 64k simplices a power law with
$\alpha = 1/4$ seems to better accommodate the data.

The data of \cite{CaKoRe94,BaSm,AmJu} suggested that one required more data
at larger volumes to decide whether the data favors a logarithmic or a
power law fit.  We have been able to get reliable data up to a volume of
128k \cite{BrMa95}.  (For the reader unfamiliar with the computationable
effort required, for the largest volume we used 30000 sweeps that took four
months on a shared IBM/RISC workstation.) Since there is not enough data to
determine $\alpha$ reliably, setting $\alpha = 1/4$ serves well enough to
distinguish the exponential from the factorial fit.

The result is that when superimposing our new points
to the fits of ref.\ \cite{AmJu} they fall very well on the power fit (quite
far indeed from the logarithmic divergence prediction) obtained from
smaller triangulations.

We have fitted our data for $k_2 = 0$ with the two forms (\ref{powerfit})
and (\ref{logfit}), by setting the power $\alpha=\frac14$.  They are both
two parameter fits.  Figure (\ref{figk4cN4}) is quite eloquent about the
success of the two fits.  The result is
\beqa
        k_4^{c(log)} & = & 0.864 + .0277 \ln N_4\ ,
\\
        k_4^{c(power)} & = & 1.252 - 1.317 N_4^{-\frac14}\ .
\eeqa
The power fit has a value of the $\chi^2 $
which is ten times better than the logarithmic one. We have also
tried $3$ parameter fits. In the power fit we have left the power as a free
parameter, while in the logarithmic fit we have added a volume scale term
$N_4^0$, as in $\ln(N_4-N_4^0)$.  Both fits improve quite a lot, but the
power fit stays far superior to the logarithmic fit (the $\chi^2$ ratio is
now $3$). While such a power fit (where the best power is now $.36 \pm
.04$) matches perfectly the data points, the logarithmic fit is still not
totally congruent to the data (we get $N_4^0$ of order $3000$, that is a
reasonable scale for the transient behavior). We are not very confident in
playing with many parameters, since the allowed corrections are of many
different functional forms, and it is clear that with $6$ data points they
cannot be distinguished. We just take the results of the $3$ parameter fits
as further evidence that the power fit is superior to the logarithmic
fit. Let us also note that indeed the best preferred power is surely not too
small.

What about the consistency of the numerical data? The first observation
about the data in figure (\ref{figk4cN4}) should really be that there is a
remarkable agreement in the data from four independent computer
implementations considering that the underlying algorithms are somewhat
similar but not identical.  In fact, notice that even the data from
\cite{CaKoRe94} that lead to the claim about the absence of an exponential
bound curves away from a straight line in the same way the other data sets
do.

The conclusion we draw is that {\it the fits of the numerical data largely
favor the existence of an exponential bound at $k_2 = 0$ over the
presence of a factorial bound.}

Having analyzed in detail the situation for $k_2 = 0$, we now turn to {\it
generic} values of the coupling $k_2$. In theory, the existence of an
exponential bound for any one value of $k_2\ge0$ implies existence for all
the others. But as is well known, but has not been discussed in detail in
this context, there is an important practical difference between the phases
for $k_2$ below and above the critical value $k_2^c \approx 1.1$. For large
positive $k_2$ the simplicial complex is in an elongated phase with an
intrinsic dimension close to two, while for negative $k_2$ the intrinsic
dimension diverges to infinity and the simplicial complex becomes extremely
crumpled. One of the most intriguing and attractive features of simplicial
quantum gravity is that at $k_2^c$ the intrinsic dimension is close to four
\cite{AgMi93,BaSm2} (for simplicity we ignore here the problem of giving the
best definition of the intrinsic dimensionality of the system).

The point is that the two phases are not only different, but there is a
genuine asymmetry.  Note that at $k_2^c$ the intrinsic system size for $N_4
= 10,000$ is of the order of $(10,000)^{1/4} = 10$, while at $k_2 = 0$ it
is $(10,000)^{1/10}\approx 2.5$. Therefore, what constitutes a {\em large
volume} that guarantees the absence of finite size effects depends very
sensitively on the value of $k_2$ \cite{Br93}. For example, the asymmetry in
the susceptibility present in these systems may be due to such effects.

\begin{figure}
\epsfxsize=350pt
\centerline{\epsffile{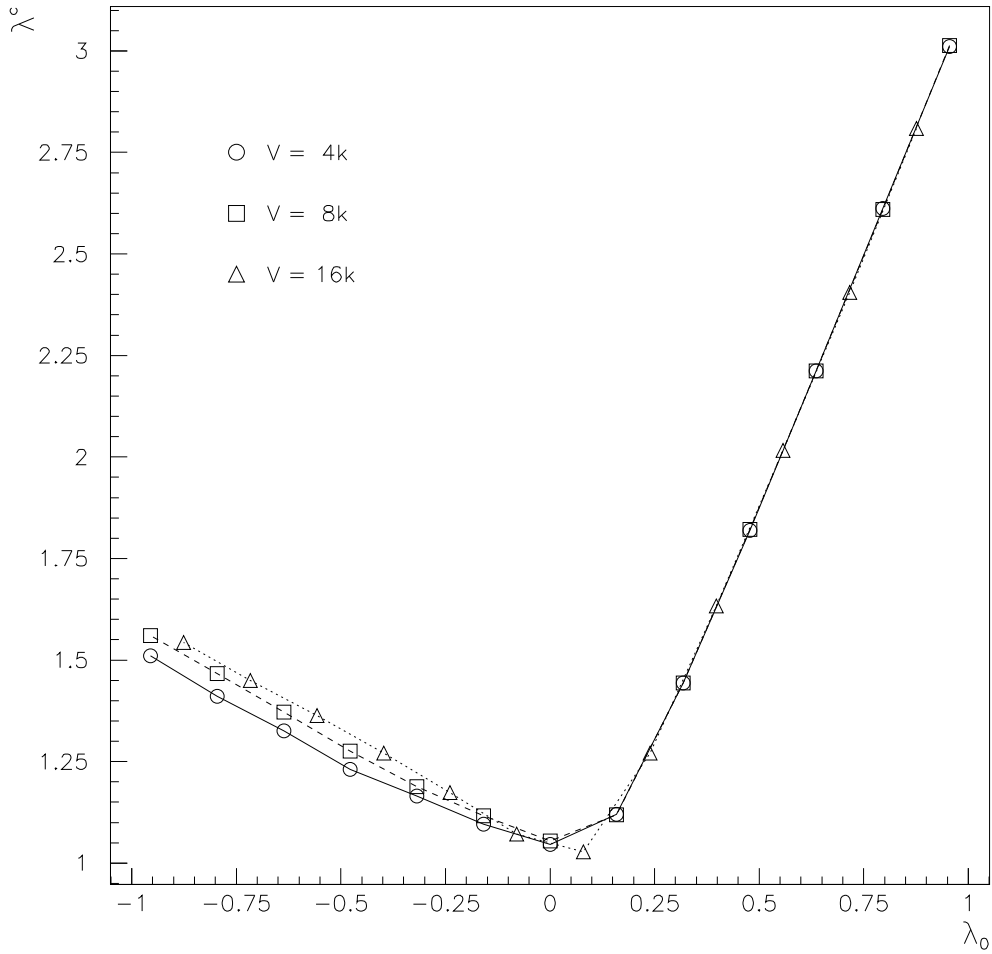}}
\caption[a]{\protect\label{figlcl0}
$\lambda^c$ versus $\lambda_0$. Indications of a phase transition are
found near $\lambda_0 = 0.18$.}
\end{figure}

With regard to the discussion of the exponential bound one should
therefore consider the whole $k_2$ range.  Such data already exist in
\cite{AgMi92,Br93} and were improved upon near the transition in
\cite{BaSm} but were not considered in \cite{CaKoRe94,AmJu}.  For
concreteness we show in figure (\ref{figlcl0}) a plot of
$\lambda^c(k_2)$ versus $\lambda_0\sim 1/G$ for $N_4=4k,8k,16k$ based
on \cite{Br93}, which for our purpose is better suited than the more
accurate data of \cite{BaSm} since figure (\ref{figlcl0}) extends to
extreme values of $k_2$.  The constants are defined by the relations

\beq
  k_2 = 2 \pi \lambda_0\ ,
  \quad  k_4 = \lambda + 10 \alpha \lambda_0\ .
\eeq
There is a definite volume dependence for $k_2 < k_2^c$ while above the
transition no volume effect is discernible.  The linear transformation from
$k_2$ and $k_4$ to the cosmological constant $\lambda$ is useful for
magnifying the volume dependence which is invisible in this range of
coupling constants for $k_4^c(k_2)$ \cite{Br93}.  This is discussed in
\cite{BaSm}, but even when explicitly looking for a small volume dependence
for $k_2$ clearly above $k_2^c$, none is found.  In this region the plot
analogous to figure (\ref{figk4cN4}) appears to be a perfectly horizontal
straight line, i.e.\ there are no detectable polynomial corrections to the
exponential bound.

The discussion can be taken one step further by noticing that the critical
value $k_2^c$ of $k_2$ moves to larger values with increasing volumes
\cite{AgMi93,BaSm}. For larger volumes at $k_2 = 0$ the finite size effects
become even more pronounced (internal dimension up to 50).  Given that for
extreme values of $k_2$ the simplicial complex freezes and $k_4^c(k_2)$
becomes a perfect straight line with different slopes, the shift in $k_2^c$
keeping the part $k_2 > k_2^c$ in figure (\ref{figlcl0}) fixed translates
directly into (part of) the volume dependence in the range $k_2 < k_2^c$.
When this effect constitutes the significant part of the volume dependence
(for large enough volume), then the volume dependence of the critical value
$k_2^c$ can be estimated by the volume dependence of $k_4^c$ for a small
enough but fixed value of $k_2$.  In particular, if there is no exponential
bound, then $k^c_2 \rightarrow \infty$ with $N_4\rightarrow \infty$.

It is instructive to examine the condition for the critical line in the
Monte Carlo simulations (here we follow \cite{Br93,BrMa93}). Consider the
ergodic random walk in the space of triangulations of $S^4$ consisting of
the five standard moves, where for the move of type $i$ an $i$-simplex is
replaced by a $(4-i)$-simplex. On the critical line, the average volume
$N_4$ is constant, and therefore $N_2$ must also be constant since it is
bounded.  This means that the average variations $\overline{\delta N_j}$
must vanish,
\beq
   \overline{\delta N_j} \sim \sum_{i=0}^4 \Delta N_j(i) p_i = 0,
\eeq
where $p_i$ is the probability with which a move of type $i$ is performed
on average, and $\Delta N_j(i)$ is the change in $N_j$ due to that
move. Since the moves are independent, we obtain
\beq
  p_0 = p_4,  \quad p_1 = p_3,
\label{pppp}
\eeq
on the critical line.

Since the action is linear in $N_2$ and $N_4$, and since the moves are
local, we can be more specific about the conditions on the $p_i$. The $p_i$
can be chosen to be
\beq
        p_i = [e^{-\Delta S(i)}] \, p_i^{geo}\ .
\eeq
The bracket is the Metropolis weight, its key feature being that it depends
only on the type of move and not on the $N_i$ or the triangulation in
general. While the action looks quite trivial, all the non-trivialities
are hidden in the probability $p_i^{geo}$ for a move to be allowed by the
geometric constraints on the triangulation. (Detailed balance is
incorporated in the way the moves are chosen. Potentially, there is a
factor of order $O(1/N_4)$.)

Therefore (\ref{pppp}) is equivalent to
\beqa
        k_4^c &=& \frac{5}{2} k_2 - \ln p_0^{geo},
\label{k1} \\
        k_4^c &=& 2 k_2 - \ln \frac{p_1^{geo}}{p_3^{geo}},
\label{k2}
\eeqa
where we have used that $p_4^{geo} = 1$.
The question of the existence of an exponential bound has therefore been
translated into the question whether there exist appropriate bounds on
the $p_i\equiv p_i(k_4,k_2,N_4)$ which are independent of $N_4$.

First of all, $p_0 \le 1$ implies that $k_4^c$ is bounded from below by
$2.5 k_2$. The hard part is to find a suitable lower bound on $p_0$, for
example, and although it may be possible to do so by some more detailed
analysis of the space of triangulations, we do not have a conclusive
argument. Notice that since moves of type 4 are always allowed, we have
that $p_0 > 0$. However, a naive counting of possible moves of type 0 and
4 around a fixed background triangulation gives $p_0 \sim 1/N_4$, which
would be the divergent scenario, but the same kind of counting would also
make 2d divergent. The counting is, of course, difficult because moves of
type 1, 2, and 3 may change the geometric constraints.

Coming from the numerical side, it is quite suggestive that e.g.\ the data
for $N_4=4k$ in figure (\ref{figlcl0}) corresponds to $k_4^c(k_2 \ge 4.0) =
2.497 k_2 + c$ and $k_4^c(k_2 \le -4.0) = 2.002k_2 + d$.  This means that
in the extreme $k_2$ regions the relevant geometric probabilities must be
independent of $k_2$ (combining (\ref{k1}) with (\ref{k2}) gives a factor
of $\exp (k_2/2)$ for the opposite side).

Considering the general structure of the phase diagram, the volume
dependence can also be understood on the level of the random walk as
follows. It is the moves of type 4 that drive the system into the elongated
phase ($\Delta N_2(4)/\Delta N_4(4) = 2.5$), while moves of type 1 drive to
the crumpled phase ($\Delta N_2(3)/\Delta N_4(3) = 2.0$).  Depending on
$k_2$ the random walk is driven towards one of the bounds in
$2<N_2/N_4<4$. One of the two possible phases, the elongated phase, is
therefore characterized by low order vertices, and the average order does
not depend on $N_4$ since a maximal elongation can be obtained for a rather
small number of simplices. Hence $p_0^{geo}$, which is the ratio of the
number of vertices of minimal order to the number of all vertices, is
expected to be independent of $N_4$ in the elongated phase. On the other
hand, in the crumpled region the average order of vertices is driven
towards large values, and the average order will grow with
$N_4$. Hence $p_0^{geo}$, which is defined by the low order tail of the
vertex order distribution,  goes to zero with $N_4$ in the crumpled phase.
Equation (\ref{k1}) gives the corresponding volume dependence of $k_4^c$.

A very attractive possibility is that sophisticated methods to count
triangulations like \cite{BaBrCaMa} may also allow to estimate the basic
probabilities $p_i^{geo}$. The necessary extension is to counting
triangulations subject to constraints, for example that there is
a certain number of $i$-simplices of minimal order that therefore can be
removed. If this can be done, there could be analytical predictions for
the Monte Carlo random walk.

In conclusion, when looking for evidence for an exponential bound in the
numerical data of simplicial quantum gravity in four dimensions, one should
take the whole range of $k_2$ into account. If one insists on looking in
the crumpled phase at $k_2=0$, the numerical data strongly support the
validity of an exponential bound.

\section{Is there a measure ambiguity?}

In the transition from the continuum path integral to the triangularized
model in (\ref{conttodisc}) we tacitly assumed that each triangulation
carries equal weight.  Strictly speaking there should be a symmetry factor
which is conventionally ignored since it presumably has a negligible
effect.  What we want to stress in this section is that the assumption of
uniform weight (modulo symmetries) is quite a serious matter since it is
completely open whether there are different universality classes for
simplicial quantum gravity in four dimensions.  In fact, we want to discuss
some data obtained for a one-parameter family of non-uniform measures that
seem to show such non-universality.

We have selected not one but a family of measures in order to
investigate the influence of the measure in a rather general setting.
Our choice is guided by diffeomorphism invariance of the measure
\cite{MEASURE} but ignores more sophisticated arguments like BRST
invariance. We have studied, as a function of $n$, a measure
contribution of the form
\begin{equation}
  \prod_x g^{n/2}\ ,
\end{equation}
i.e. in the triangulated theory $S_E[T]$ is replaced by $S[T] = S_E[T]
+ S_M[T]$, where
\begin{equation}
  S_M = - n \sum_a \log \frac{o(a)}{5}\ .
\end{equation}
The sum runs over all $0$-simplices (sites) of the manifold, and
$o(a)$ is the number of $4$-simplices which include the site $a$. We
considered $n$ in the interval from $-5$ to $5$. The case $n=0$
repeats simulations with the trivial, uniform measure, which can be
compared with previous results.

In two-dimensional dynamical triangulations, it has been already observed
that on finite lattices, as expected, the phase diagram depends on the
measure coupling (e.g.  \cite{DaJuKrPe87}).  However, in this case the
measure term is naively an irrelevant operator, and once an extrinsic
curvature term has been added it has been shown that the phase diagram
\cite{BoCoHaHaMa} is insensitive to this term in the large lattice limit
(also at the crossover point).

In the 4d case the phase diagram is also expected to depend on $n$ (if
one sends $n$ to $\pm\infty$, then the measure term would dominate the
action). It is an interesting question to ask whether for some reasonable
values of $n$ one obtains different universality classes. The conventional
choice of uniform measure is valid if this is not the case.

Let us summarize our results. We find that the measure factor plays an
important role, and that the critical behavior does depend on $n$.
Varying $n$ not only changes non-universal quantities like the value of
the critical coupling, but there are indications that it changes the
actual critical behavior.

\begin{figure}[t]
\epsfxsize=350pt
\centerline{\epsffile{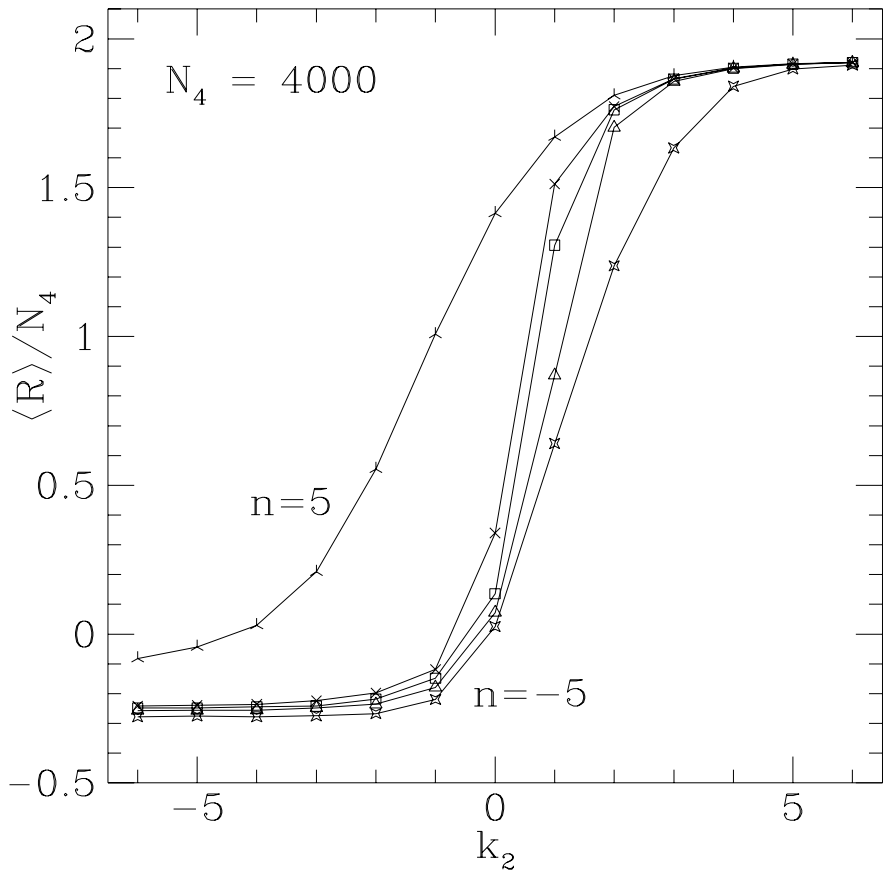}}
\caption{Expectation value of $R/V$ for
$n=-5, -1, 0, 1, 5$. The curves are ordered from the smallest $n$ at the
bottom upwards.}
\protect\label{figRk2}
\end{figure}

\begin{figure}[t]
\epsfxsize=350pt
\centerline{\epsffile{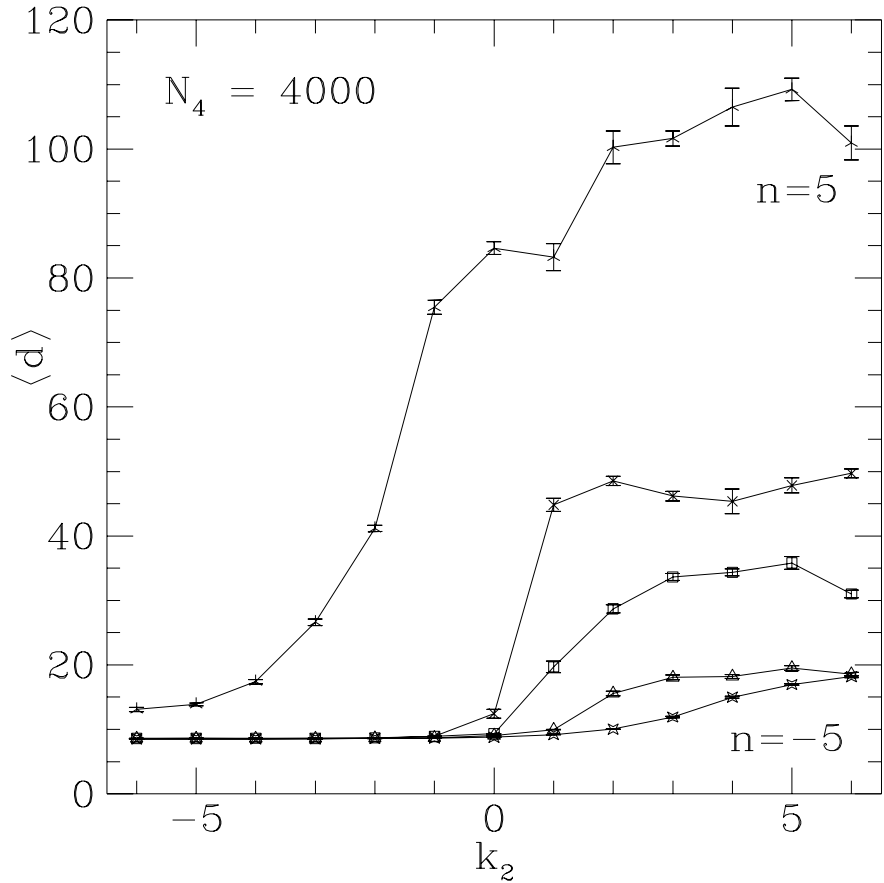}}
\caption{Expectation value of $d$ for $n=-5, -1, 0, 1, 5$, $N_4=4000$; curves
ordered as in figure (\protect\ref{figRk2}).}
\protect\label{figdk2}
\end{figure}

In figure (\ref{figRk2}) we plot the average curvature $R/V$ for $V\equiv
N_4=4000$ as a function of the coupling $k_2$ for different values of $n$,
$n=-5$ for the lowest curve, then $n=-1$, $0$, $1$ and $n=5$ for the upper
curve.  In figure (\ref{figdk2}) we plot the average distance (in the
internal space) of two 4-simplices.  We count the minimum number of steps
from 4-simplex to 4-simplex across 3-simplex faces that connect a pair of
4-simplices and average over all 4-simplices and random manifolds.

Both figures show that the measure operator has a pronounced effect.
Increasing the coupling of the measure term leads to a continuous,
monotonous deformation of the curves. Notice that the curves are not
just shifted. In the case of $R/V$, the singularity seems stronger for
$n \simeq 0$, where the jump in $R/V$ is quite sharp. The distance $d$
has a sharper jump for $n=1$, where it seems to jump from one constant
value to another constant. Smaller values of $n$ show a slower
increase in $d$.

For large absolute values of $n$, especially for $n=-5$, the plots
show a weaker singularity. The profile of $R/V$ hints less at a sharp
jump than the former cases, and the distance increases very smoothly
from a critical value of $k_2$, $k_2^c(n)$, on. When $n$ increases to
the value of $5$ the system seems to loose criticality on an absolute
scale.  Its behavior through the crossover is quite smooth.

A critical value of $k_2^c$ can be defined, for example, as the point where
the distance value starts to change. But for the $n=5$ case the transition
point is not very clear. Let us note that such a value of $k_2^c$ changes its
sign as a function of $n$.

Our conclusion is that the measure term has a strong effect, which
seems difficult to reabsorb in a simple renormalization of the
critical coupling. Always keeping in mind that a precise finite size
study is required before making quantitative statements,
we believe there are two basic possibilities. The first possibility is
that there is only one universality class, and that all the theories
we have studied do asymptotically show the same critical behavior. In
this case the rate of approach to the continuum limit is strongly
influenced by $n$. We will select the theory with faster convergence
to the continuum.

The second possibility (which is the most interesting one) is that the
measure factor changes the universality class. Our results, albeit
preliminary, seem to hint in this direction. In this case we could
have a critical value of $n_c$, and transitions belonging to different
universality classes. This is a very appealing scenario, and here the
lattice theory could make its own original contribution. It
could be possible to pick out the correct measure, on the lattice, by
requiring a particular expectation value and scaling behavior of some
physical observable. Such a prescription would be a powerful tool,
turning the discrete version of the theory from a source of
indetermination into a completely determined scheme.

\section{Conclusions}

It is not easy to summarize such an evolving scenario. Things look
good, and interesting.

The number of triangulation does not seem to increase in a
pathological way, and the exponential bound seems to be valid. This is
the first evidence that makes our hope of finding interesting
phenomena stronger.

Different models based on different choices of the lattice measure have
quite different behaviors. One will have to investigate in more detail if
all the lattice theories will have the same continuum behavior or if we are
finding a more complex phase diagram.

Finally, the new results of \cite{continuous} seem to fortify the hope that
a critical theory could be generated at one point of the phase diagram.
Simplicial quantum gravity looks like a promising field, definitely worth
of further investigations.


\newcommand{\bib}[1]{\bibitem{#1}}

\newcommand{\apny}[1]{{\em Ann.\ Phys.\ (N.Y.) }{\bf #1}}
\newcommand{\cjm}[1]{{\em Canadian\ J.\ Math.\ }{\bf #1}}
\newcommand{\cmp}[1]{{\em Commun.\ Math.\ Phys.\ }{\bf #1}}
\newcommand{\cqg}[1]{{\em Class.\ Quan.\ Grav.\ }{\bf #1}}
\newcommand{\grg}[1]{{\em Gen.\ Rel.\ Grav.\ }{\bf #1}}
\newcommand{\jgp}[1]{{\em J. Geom.\ Phys.\ }{\bf #1}}
\newcommand{\ijmp}[1]{{\em Int.\ J. Mod.\ Phys.\ }{\bf #1}}
\newcommand{\jmp}[1]{{\em J. Math.\ Phys.\ }{\bf #1}}
\newcommand{\mpl}[1]{{\em Mod.\ Phys.\ Lett.\ }{\bf #1}}
\newcommand{\np}[1]{{\em Nucl.\ Phys.\ }{\bf #1}}
\newcommand{\pl}[1]{{\em Phys.\ Lett.\ }{\bf #1}}
\newcommand{\pr}[1]{{\em Phys.\ Rev.\ }{\bf #1}}
\newcommand{\prl}[1]{{\em Phys.\ Rev.\ Lett.\ }{\bf #1}}

\end{document}